%% file: omnet2014.tex
\documentclass[conference]{IEEEtran}

\usepackage[utf8]{inputenc}
\usepackage{paralist}
\usepackage[hang]{subfigure}
\usepackage{multirow}
\usepackage{array} % for 'm' alignment in tabular (obviously ;) )
\usepackage{units}
\usepackage{textcomp}
\usepackage{comment}
\usepackage{relsize}
\usepackage{graphicx}

\DeclareFontFamily{U}{aer}{}
\DeclareFontShape{U}{aer}{m}{n}{<-6>eurm5<6-8>eurm7<8->eurm10}{}
\DeclareSymbolFont{myMuFont}{U}{aer}{m}{n}
\DeclareMathSymbol{\mymuSym}{\mathord}{myMuFont}{"16}
\newcommand{\mymu}{\ensuremath{\mymuSym}}

\newcommand{\comsysmarker}{\IEEEauthorrefmark{1}}
\newcommand{\kthmarker}{\IEEEauthorrefmark{3}}

\newcommand{\fig}[1]{Figure~\ref{#1}}

\newcommand{\ie}{i.$\,$e.,}
\newcommand{\eg}{e.$\,$g.,}
\newcommand{\etal}{et$\,$al.}
\newcommand{\omnet}{OMNeT\raise.1ex\hbox{++}}
\newcommand{\ns}{\mbox{ns-3}}
\newcommand{\us}[1]{\unit[#1]{\mymu}s}
\newcommand{\pc}[1]{\unit[#1]\%}

\begin{document}

\title{Enabling Distributed Simulation\\of OMNeT\raise.2ex\hbox{++} INET Models}

\author{
\IEEEauthorblockN{Mirko Stoffers\comsysmarker, Ralf Bettermann\comsysmarker, James Gross\kthmarker, Klaus Wehrle\comsysmarker}
\IEEEauthorblockA{\comsysmarker Communication and Distributed Systems, RWTH Aachen University}
\IEEEauthorblockA{\kthmarker School of Electrical Engineering, KTH Royal Institute of Technology}
\IEEEauthorblockA{stoffers@comsys.rwth-aachen.de}
}

\maketitle
\begin{abstract}
	Parallel and distributed simulation have been extensively researched for a long time.
	Nevertheless, many simulation models are still executed sequentially.
	We attribute this to the fact that many of those models are simply not capable of being executed in parallel since they violate particular constraints.
	In this paper, we analyze the INET model suite, which enables network simulation in \omnet\, with regard to parallelizability.
	We uncovered several issues preventing parallel execution of INET models.
	We analyzed those issues and developed solutions allowing INET models to be run in parallel.
	A case study shows the feasibility of our approach.
	Though there are parts of the model suite that we didn't investigate yet and the performance can still be improved, the results show parallelization speedup for most configurations.
	The source code of our implementation is available through our web site at {\tt code.comsys.rwth-aachen.de}.
\end{abstract}

\section{Introduction}
\label{intro}

Parallel and distributed simulation has been a major research topic for a long time.
A lot of effort went into the research and development of parallel simulation engines.
However, although sophisticated scheduling and synchronization techniques are available, many simulation models are still executed sequentially.
This results in long execution times and defers the development process of new products.

Unfortunately, not every simulation model can directly benefit from the manifold research advances in the area of parallel simulation frameworks.
Those simulation models are not developed for parallel execution, but are written with only sequential simulation in mind.
Hence, they are simply not capable of parallel execution.
This situation motivates the investigation of existing simulation models in order to determine the causes preventing parallel execution.
The results of such an analysis can be used to develop solutions that might also be applicable to other simulation models suffering from similar problems.

In this work we investigate the INET model suite \cite{inet} for \omnet\ \cite{varga2001} as a widely used suite for network simulation with regard to its parallelizability.
In particular we make the following contributions:
\begin{inparaenum}
	\item We identify the issues which render parallel execution of INET based simulation models impossible.
	\item We describe our solutions to allow parallel execution of broad parts of the INET model suite.
\end{inparaenum}

The results of our investigations can be applied by model developers facing similar problems with their simulation models.
The source code of our modifications is publicly available\footnote{The source code with our modifications is available at our project side: https://code.comsys.rwth-aachen.de/redmine/projects/parallel-inet}.

The remainder of this paper is structured as follows.
After providing an overview of the state of the art, we describe the results of our analysis and the solutions to the parallelization problems.
After that, we describe the results of a case study performing a parallel simulation of a 4,000 node network.
Finally, we conclude with a summary and outlook on future efforts.

\section{Background and Related Work}
\label{back}

Discrete Event Simulation (DES) has been used for a long time for simulating computer networks.
Today, \ns\ \cite{ns3} and \omnet\ \cite{varga2001} are the most common open-source tools used for network simulation.
While \ns\ is specifically tailored for this purpose, \omnet\ is a general purpose DES framework, and the INET model suite \cite{inet} brings network simulation capabilities to \omnet.
Both frameworks are by themselves capable of parallel simulation.

However, since the common components of a network stack are directly integrated into the \ns\ simulator, they are, like \ns\ itself, directly designed for parallel execution.
Hence, \ns\ shows promising results in parallel network simulation \cite{pelkey2011}.
On the other hand, the INET suite for \omnet\ was designed for sequential simulation without special care for fitness for parallel simulation.
To this end, the releases of INET cannot be executed in parallel without additional effort.

Efforts to parallelize the INET suite have been taken by Nowak and Nowak \cite{nowak2007}.
However, the evaluation results show that parallelization speedups are only achieved for extraordinary long link delays in the area of hundreds of milliseconds.
Furthermore, the authors concentrated only on TCP and IP, omitting essential protocols like, \eg\ Ethernet such that no realistic MAC layer protocol can be used.
INET includes powerful auto-configuration tools to ease the setup of complex simulation scenarios, such that the network size can be increased with a negligible amount of manual effort.
Unfortunately, the approach by Nowak and Nowak does not target these auto-configuration tools.
We argue that parallel simulation is particularly interesting for large scenarios, hence comfortable methods to set up the networks are as important as a broad coverage of the most important protocols.

For this reason, we analyzed the issues that prevent models using the INET suite from running in parallel and designed a parallelized version which supports protocols on the entire network stack as well as comfortable setup tools for large networks.

\section{Analysis and Design}
\label{design}

Sekercioglu \etal\ state in \cite{sekercioglu2003} the requirements for simulation models to be run in parallel with \omnet.
The authors state that global variables cannot be used to exchange information between LPs.
Directly calling methods on remote LPs is not allowed, and direct sending of messages to modules on different LPs is only partially supported.
Dynamic changes to the network topologies during runtime are not yet supported.
For efficient parallelization, sufficient lookahead needs to be present by means of transmission delays on the links between modules on different LPs.
We analyzed the INET model suite with regard to those constraints.

Our observations revealed that in fact, during a running simulation, most of these constraints are met by the common protocols Ethernet, IP, TCP, and UDP.
However, during initialization and configuration the constraints are violated in several cases.

For this reason, we further investigated the issues during setup and developed a method for distributed initialization of simulation models, described in the following.
After that, we describe the additional steps required to run INET models yielding correct results.

\subsection{Distributed Multi-Stage Initialization}

Our analysis of the INET model suite encountered three major parallelization issues during intialization:
\begin{compactenum}
	\item The INET model suite provides a comfortable way of assigning MAC addresses automatically to all Ethernet devices in a simulation model.
		During this auto-assignment procedure, however, the next available MAC address is stored in a global variable.
		In a distributed simulation, independent copies of these global variables exist.
		Hence, two LPs assign equal MAC addresses to different network interfaces.
	\item To decide whether it is necessary to perform certain operations on network devices, Ethernet modules need to determine whether they are connected to a remote Ethernet device.
		In INET, the connection state is determined by invoking the {\tt isConnected} function on the local gate and on all gates along the path to the remote end.
		This works well as long as the remote node resides on the same LP.
		However, as soon as the remote end resides on a different LP, the function can no longer be called on the correct object, and the connection state is determined incorrectly.
	\item INET provides a configurator module to assign IP addresses to the network nodes as specified by the user.
		This powerful configurator allows to assign, for example, several nodes IP addresses from a given subnet or even to automatically assign network addresses to the subnets.
		For this purpose, the IPv4 configurator searches for available network interfaces by invoking functions on all nodes in the simulation.
		However, since the configurator is an \omnet\ module, it needs to be assigned to an LP.
		Hence, for all network nodes on different LPs the configurator cannot find the interfaces.
		To this end, no the interface on the remote LPs are not set up at all.
		Instantiating additional configurators on the remote LPs allows assignment of IP addresses to all nodes, but results in equal IP addresses being assigned to different nodes.
\end{compactenum}

A prevailing goal of our work was to enable distributed simulation of INET based simulation models without requiring to waive comfortable options like auto-assignment of MAC addresses or the features of the powerful IPv4 configurator.
Hence, we developed a method that enables distributed initialization of INET models without structural changes to the simulation model.
However, the structure of the existing INET components like the IPv4 configurator requires exchanging information during the intialization phase.
The only method for exchanging information in a distributed execution of \omnet\ models is to schedule events on remote LPs.
However, this option is not feasible for initialization since the events will not be executed before the start of the simulation.

\begin{figure}
	\centering\includegraphics[width=0.95\columnwidth]{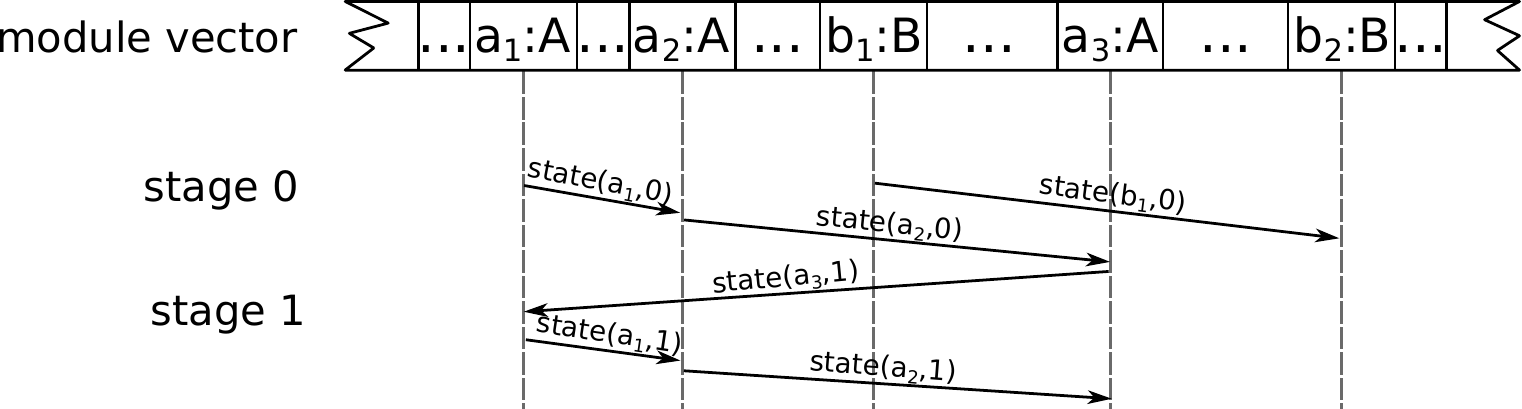}
	\caption{Distributed Multi-Stage Intialization:
		The classes A and B are registered for DMSI, hence their instances exchange state information respectively.
		In stage 0, the state of $a_1$ is transmitted to $a_2$, and the modified state is forwarded to $a_3$.
		After that, the modified state is sent to $a_1$ which continues with the initialization of stage 1.
	}
	\label{fig:dmsi}
\end{figure}

Our solution to this issue is the development of an extension for the simulation framework, called {\em Distributed Multi-Stage Intialization} (DMSI, see \fig{fig:dmsi}).
The purpose of DMSI is to allow multi-stage initialization as supported by \omnet\ with additional support for the LPs to exchange information between two stages.
To use DMSI with INET, we implemented it in \omnet.

To this end, our implementation of DMSI for \omnet\ provides a macro allowing classes to register for DMSI.
Each instance of a registered class is then initialized one at a time.
The multi-stage initialization feature of \omnet\ is still provided, \ie\ we perform stage $n$ only after all modules on all LPs have completed stage $n-1$.
The first module that is initialized in stage 0 needs to create a serializable state object which the DMSI core passes to the next module.
This module can then read and modify that state, and the modified state is passed on.
In stage $n>0$ the first module receives the state returned by the last module in stage $n-1$.
Hence, the state can be used to encapsulate global variables or other information to be shared during initialization.

In general, their are two basic use cases to map non-parallelizable initialization concepts of sequential simulation to DMSI:
{\em Global variables} can be encapsulated in the state, allowing every module to access the variable.
{\em Direct method calls} can be mapped to a request-response scheme.
In stage $i$, the caller of the method would enqueue a request to call the method and the required input data to the shared state, such that the state contains a list of all methods to be called.
In stage $i+1$, the methods are executed on the target LPs, and the output data is enqueued to the state.
In stage $i+2$, the callers can access the results and proceed with their initialization.

We used DMSI to solve the three initialization problems mentioned above:
\begin{compactenum}
	\item We register the class {\tt EtherMAC} for DMSI and use the state to store the last assigned MAC address
		To this end, we have moved the global variable to the DMSI state such that every module can access the sole instance of the variable.
	\item Ethernet modules requesting connection states can do so in stage 0 by enqueuing their request to the state.
		In the next stage, the requested Ethernet modules enqueue their answer to the state, hence modules can use this information from stage 2 on.
	\item The only structural modification needs to be performed for the IPv4 configurator.
		Assigning the configurator to an LP and allowing it to configure remote nodes would result in a huge amount of data to be stored in the state, such that the nodes can be configured correctly.
		Hence, parallel INET models need to install an IPv4 configurator on each LP.
		Since the configurators nevertheless need to know the topology of the simulated network, we register the configurator class for DMSI.
		Each configurator then enqueues all local interfaces that need to be configured to the state, hence each configurator gains global knowledge about the interfaces.
		By shifting the actual initialization to the next stage, each configurator can now access the topology from the shared state and can configure the local interfaces as in a sequential simulation.
\end{compactenum}

This allows the correct initialization of \omnet\ INET models in a distributed simulation.

\subsection{Further Issues}

Additionally to the issues solved by DMSI, our analysis uncovered problems with the assignment of UDP socket IDs.
INET requests globally unique numbers from \omnet, and casts the unsigned results to signed values.
Since \omnet\ divides the available integer space into ranges for each LP to ensure global uniqueness, in distributed simulation the unsigned-to-signed cast results in an integer overflow.
Though this is not directly a parallelization problem, it does not occur in sequential simulation.
Therefore, the problem was not encountered as long as INET was not executed in parallel.

However, the solution for this issue is simple, since for the UDP socket IDs global uniqueness is not required.
Instead, the socket needs to be uniquely identifiable only from that node.
Hence, we replaced the globally unique numbers by locally unique numbers using member variables to solve this issue.

Furthermore, serialization and deserialization functions had to be implemented to correctly transmit, for example, IP addresses to remote LPs.

The above mentioned actions allowed to run INET models in parallel and derive correct results.
However, for debugging and validation it is helpful if the parallel simulation yields exactly the same results as a sequential run.
Due to the nature how \omnet\ handles random numbers, this is not guaranteed by default.
In a sequential simulation, multiple modules can draw random numbers from the same RNG.
This is no longer possible if two modules reside on different LPs.
To yield equal results in parallel and sequential execution, we created separate RNGs for each module, assuring random numbers to be independent on the partitioning and the number of LPs.

This allowed us to validate the results and we retrieved exactly the same results in sequential and parallel execution.

\section{Evaluation}
\label{eval}

\begin{figure}
	\centering\subfigure[The Backbone Network consisting of 57 nodes.]
	{\includegraphics[width=0.95\columnwidth]{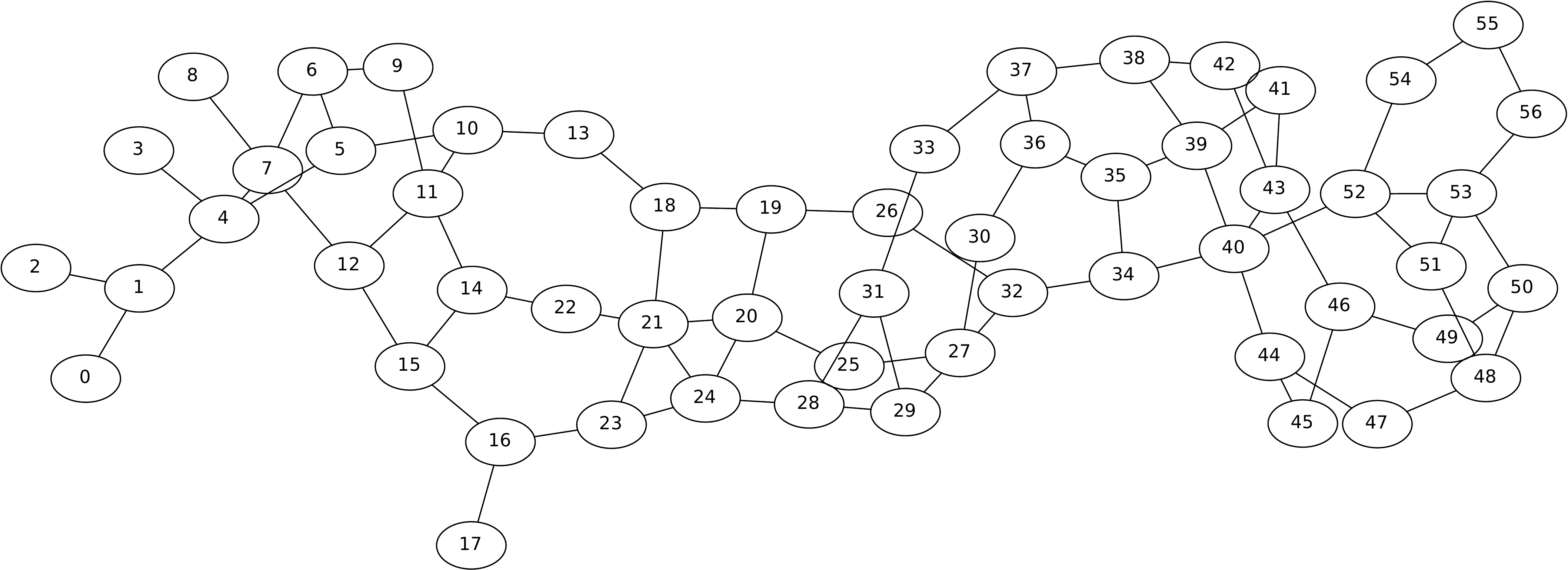}\label{fig:evalSetup:net60}}
	\centering\subfigure[A Corporate / Campus Network. Rectangles are hosts (57 in total), ellipses are routers (13 in total).]
	{\includegraphics[width=0.75\columnwidth]{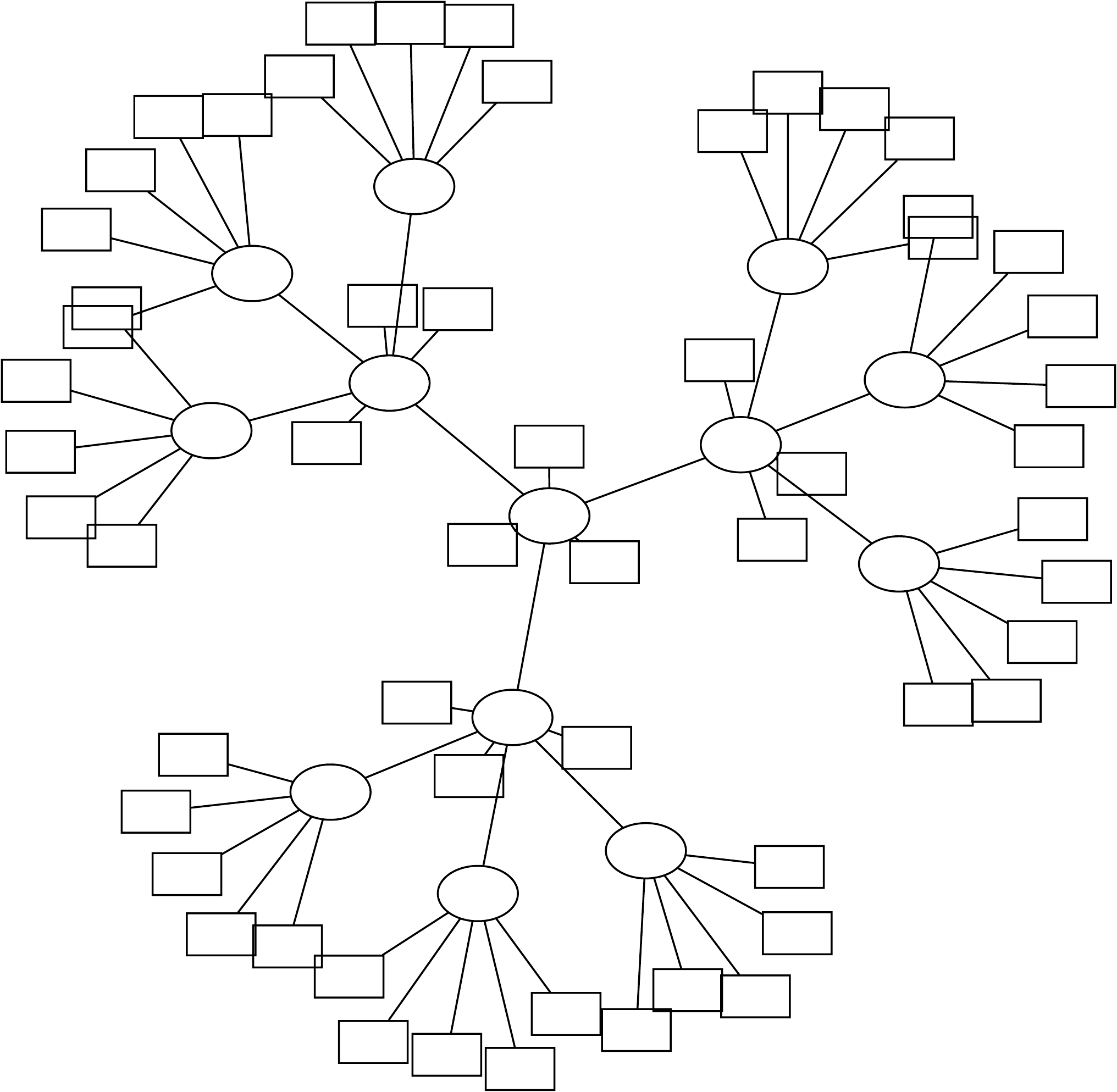}\label{fig:evalSetup:treeNet}}
	\caption{The evaluation scenario. A corporate / campus network (b) is connect to each of the routers in the backbone (a).}
	\label{fig:evalSetup}
\end{figure}

We evaluate the performance of parallel INET simulation models by means of a case study.
We use the NTT backbone topology from an \omnet\ example (see \fig{fig:evalSetup:net60}) consisting of 57 routers.
We connect a tree-structured corporate / campus network (see \fig{fig:evalSetup:treeNet}) consisting of a total of 70 nodes to each backbone router.

All connections inside the backbone and inside the corporate LANs are configured with a link delay of \us{100}.
The end hosts are connected to a router with a data rate of \unit[1]{Gb/s}, the corporate routers are connected with each other and with the backbone at \unit[10]{Gb/s}.
The routers in the backbone are connected with a data rate of \unit[100]{Gb/s}.
To investigate the influence of a wide range of different lookahead values, we vary the link delays of the connections between a LAN and the backbone from \unit[10]{ns} to \unit[5]{ms}.

All end hosts send data to random hosts via UDP using a modified version of the UDP Basic App.
To vary the amount of data transmitted to different LPs, we used two configurations:
In the first configuration, hosts select \pc{50} of the target hosts from the end hosts on the same corporate network and \pc{50} from remote networks.
In the second configuration, \pc{90} of the traffic remains local.
The hosts transmit packets with exponentially distributed inter-arrival times and exponentially distributed packet sizes.
On average, each host transmits a \unit[200]{Byte} packet every \us{20}, resulting in a moderate load of the network.

We benchmark the performance on the ``Bull MPI-S'' cluster \cite{primer2013} of the Computing Center of RWTH Aachen University.
Each node in this cluster is equipped with 12 physical cores rated at \unit[3]{GHz} and the nodes are connected by Infiniband Interconnect.
We measured the parallelization speedup for 30 repetitions with different RNG seeds and calculated the mean and the \pc{99} confidence intervals.

We used two different partitionings, using a different amount of computing nodes.
With 5 cluster nodes and a total of 58 LPs we have an LP for the backbone and an LP for each LAN.
For a benchmark on a single computing node we use 12 LPs in total.
This results in 9 LPs each maintaining 5 LANs, 2 LPs maintaining 4 LANs, and an LP with 4 LANs and the backbone.

\begin{figure}
	\centering\subfigure[\pc{50} local traffic, \ie\ \pc{50} of the packets are sent to nodes in the same corporate LAN.]
	{\includegraphics[width=0.95\columnwidth]{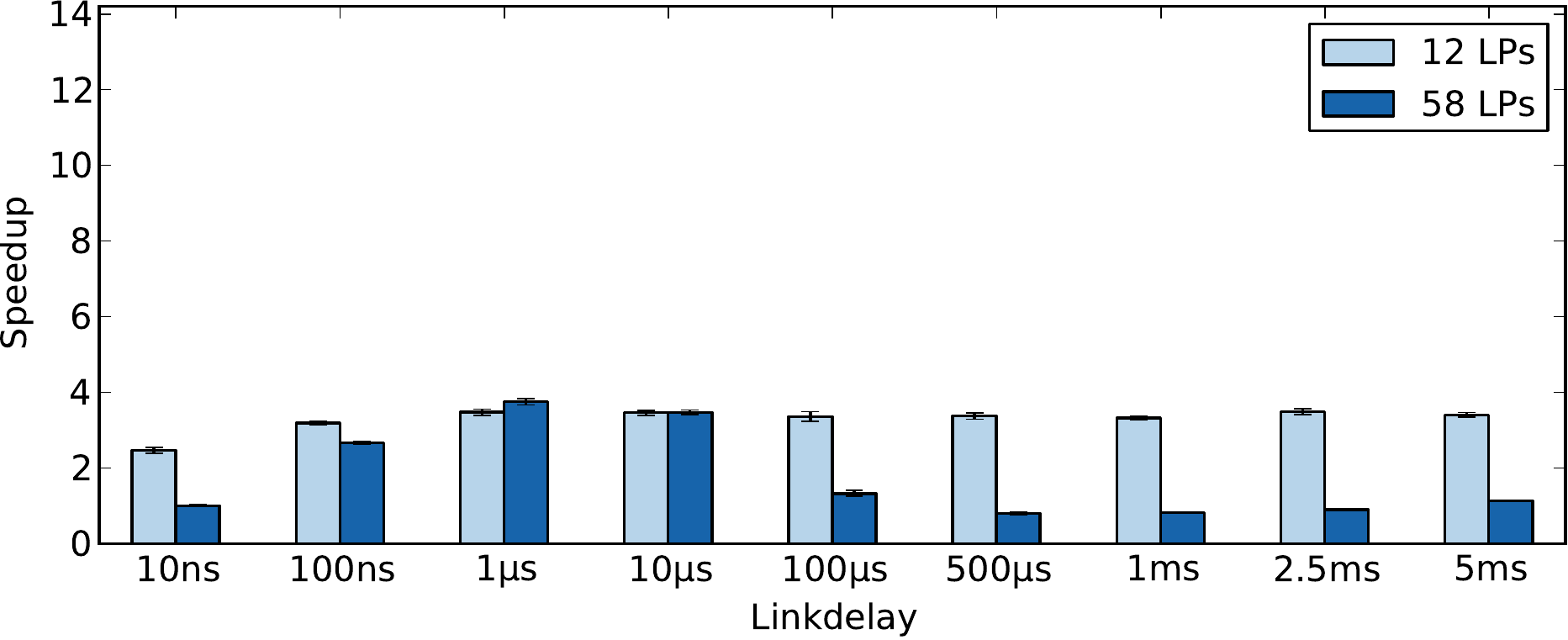}\label{fig:evalRes:fifty}}
	\centering\subfigure[\pc{90} local traffic, \ie\ \pc{90} of the packets are sent to nodes in the same corporate LAN.]
	{\includegraphics[width=0.95\columnwidth]{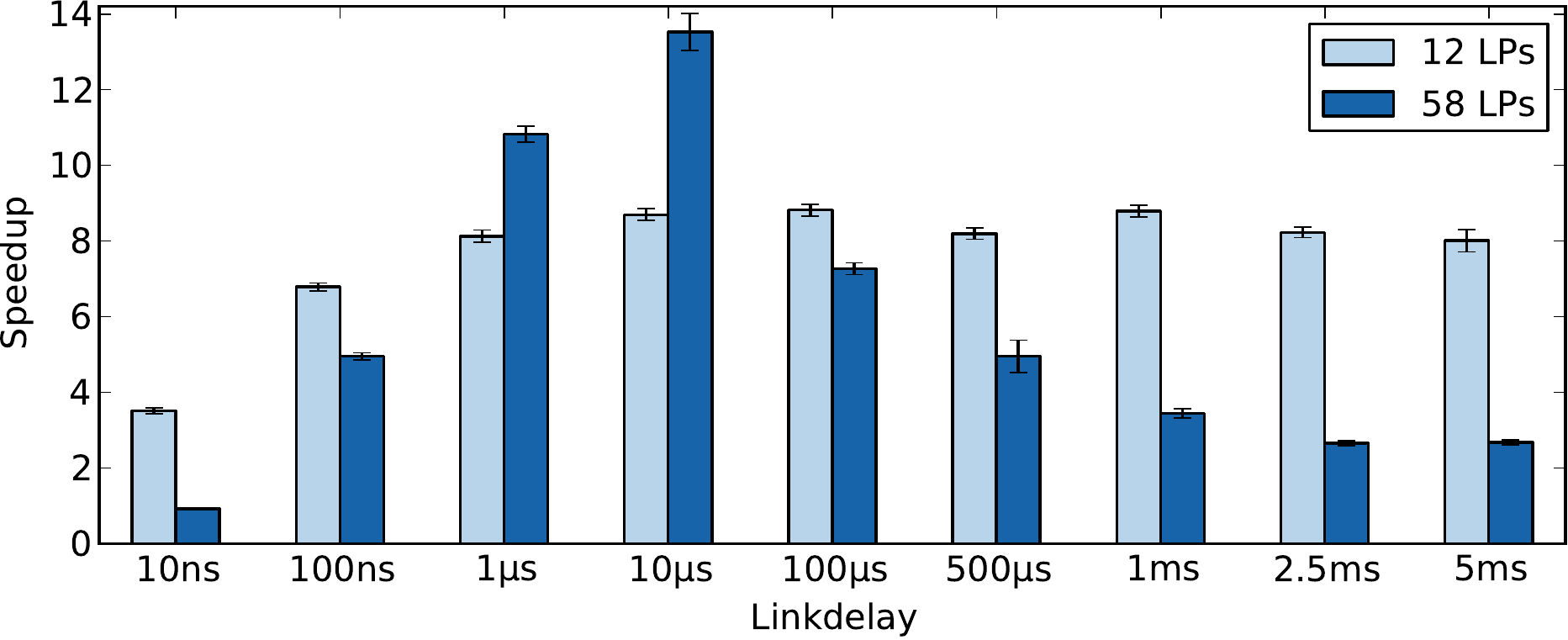}\label{fig:evalRes:ninety}}
	\caption{Average parallelization speedup of the case study with \pc{99} confidence intervals over different configurations with varying number of LPs, fraction of local traffic, and link delay.}
	\label{fig:evalRes}
\end{figure}

\fig{fig:evalRes} shows the results of the evaluation for the configurations mentioned above.

All configurations show the expected benefit when the link delay is increased from \unit[10]{ns} to \us{10}.
While a lookahead of \unit[10]{ns} results in only a little or even no speedup, for \us{10} we see a speedup of almost 14 (with \pc{90} local traffic and 58 LPs).
However, on further increase of the link delay the performance stagnates for 12 LPs and even drops for 58 LPs.
We attribute this to the fact that already a link delay of \us{10} is sufficient for parallelization, but higher link delays increase the amount of data queued at the gateways and this queue management at LP boundaries degrades performance.
This assumption is also confirmed by the fact that the performance degradation can only be observed for a simulation on multiple cluster nodes.
However, the case study shows that parallel simulation of INET models can speed up the simulation already for short link delays.

This benchmark covers the implementation of Ethernet, IPv4, ICMPv4, ARP, UDP, and the modified UDP App.
Additionally, our implementation and validation covers PPP, TCP, TCP Apps, and the HTTP App.
The most important protocols not yet analyzed and tested are WLAN and IPv6.

\section{Conclusion}
\label{conc}

In this paper, we describe our analysis of the parallelization issues in the \omnet\ model suite INET and show the feasibility of parallelizing INET with certain modifications, primarily for distributed initialization of INET models.
Our concept of Distributed Multi-Stage Initialization (DMSI) allows a simulation model to be set up for a distributed simulation run, even if the LPs require information only available at remote LPs.
We modified INET accordingly, such that it uses this concept.
To this end, models based on our modified INET suite are always set up in exactly the same way, independent of the number of LPs and independent of the chosen execution method, parallel or sequential.

Benchmarks show parallelization speedup for our case study in most configurations though potential for optimizations is still available.
Our implementation covers the most frequently used protocols like Ethernet, IPv4, TCP, and UDP as well as the adherent models like ICMPv4 or a UDP App.

Future efforts should investigate the remaining common protocols IPv6 and WLAN.
While we expect a rather straightforward procedure for IPv6, WLAN will be more challenging due to the broadcast nature of wireless channels.
Additionally, also less common protocols should be analyzed to investigate the peculiarities of those implementations.
Furthermore, improvements of the parallelization efficiency of our implementation, which have not been the focus of this work, can be performed.

\section*{Acknowledgments}
This research was funded by the DFG Cluster of Excellence on Ultra High-Speed Mobile Information and Communication (UMIC).

\input{confs.tex}

\bstctlcite{config}
\bibliographystyle{IEEEtranS}
\bibliography{literature}

\end{document}

%% file: confs.tex
% Confs:
\newcommand{\esm}[1]{Proc. of the #1 Europ. Sim. Multiconf.}
\newcommand{\ess}[1]{Proc. of the #1 Europ. Sim. Symposium}
\newcommand{\ipdps}[1]{Proc. of the #1 Intl. Parallel Distributed Processing Symposium}
\newcommand{\mascots}[1]{Proc. of the #1 Symposium on Modeling, Analysis and Sim. of Computer and Telecommunication Systems}
\newcommand{\pads}[1]{Proc. of the #1 Workshop on Parallel and Distributed Sim.}
\newcommand{\padsNew}[1]{Proc. of the #1 Workshop on Principles of Advanced and Distributed Sim.}
\newcommand{\padsSig}[1]{Proc. of the #1 ACM SIGSIM Conf. on Principles of Advanced Discrete Sim.}
\newcommand{\ppeals}[1]{Proc. of the #1 ACM SIGPLAN Conf. on Parallel Programming: Experience with Applications, Languages and Systems}
\newcommand{\simutools}[1]{Proc. of the #1 Conf. on Sim. Tools and Techniques}
\newcommand{\winter}[1]{Proc. of the #1 Winter Sim. Conf.}

% Jornals:
\def\comm{Communications of the ACM}
\def\csur{ACM Computing Surveys}
\def\sweng{IEEE Trans. on Software Engineering}
\def\pds{IEEE Trans. on Parallel and Distributed Systems}
\def\tai{Theoretical and Applied Informatics}
\def\vehic{IEEE Trans. on Vehicular Technology}